\documentclass[12pt]{iopart}

\usepackage{graphicx}
\usepackage{iopams}
\usepackage{setstack}
\usepackage[pdfborder={0 0 0}]{hyperref}

\usepackage[latin1]{inputenc}
\usepackage{dcolumn}
\usepackage{bm}

\newcommand{\ben}{\begin{enumerate}}
\newcommand{\een}{\end{enumerate}}
\newcommand{\mc}{\mathcal}

\newcommand {\BEC} {\rm{BEC}}

\newcommand {\rf} {\rm{rf}}

\newcommand{\psit}{\boldsymbol{\psi}}
\newcommand{\psip}{\psi_{\perp}}
\newcommand{\phil}{\phi_{\rm{L}}}
\newcommand{\phib}{\phi_{\rm{B}}}

\newcommand{\psil}{\psi_{\rm{L}}}
\newcommand{\psib}{\psi_{\rm{B}}}

\newcommand{\ntot}{n_{\rm{tot}}}
\newcommand{\nl}{n_{\rm{L}}}

\newcommand{\rp}{\bi{r}_{\perp}}

\newcommand{\tmu}{\tilde{\mu}}

\newcommand{\mueff}{\mu_{\rm{eff}}}
\newcommand{\geff}{g_{\rm{eff}}}
\newcommand{\omegac}{\Omega_{\rf}^{\rm{c}}}
\newcommand{\omegarf}{\delta\omega_{\rf}}

\newcommand{\Gammac}{\Gamma_{\rm{c}}}

\newcommand{\Gammasc}{\Gamma_{\rm{sc}}}

\newcommand {\BECA} {\rm{B}}
\newcommand{\phibA}{\phi_{\rm{B}}}
\newcommand{\nbA}{n_{\rm{B}}}

\newcommand{\zbar}{\bar{z}}
\newcommand{\Rzbar}{\bar{R}_{z}}

\newcommand{\dbar}{\bar{\delta}}

\begin{document}

\title{Quasicontinuous horizontally guided atom laser: coupling spectrum and flux limits}

\author{A Bernard, W Guerin, J Billy, F Jendrzejewski, P Cheinet, A Aspect, V Josse and P Bouyer}
\address{Laboratoire Charles Fabry, Institut d'Optique Graduate School, CNRS and Universit\'e Paris Sud, Campus Polytechnique, RD 128, 91127 Palaiseau, France }
\ead{vincent.josse@institutoptique.fr}

\begin{abstract}
We study in detail the flux properties of a radiofrequency outcoupled horizontally guided atom laser, following the scheme demonstrated in [Guerin W \textit{et al.} 2006 \textit{Phys. Rev. Lett.} \textbf{97} 200402]. Both the outcoupling spectrum (flux of the atom laser versus rf frequency of the outcoupler) and the flux limitations imposed to operate in the quasicontinuous regime are investigated. These aspects are studied using a quasi-1D model, whose predictions are shown to be in fair agreement with the experimental observations. This work allows us to identify the operating range of the guided atom laser and to confirm its good promises in view of studying quantum transport phenomena.
\end{abstract}

\pacs{03.75.Pp, 39.20.+q, 42.60.Jf,41.85.Ew}

\submitto{\NJP}

\maketitle

\section{\label{intro}Introduction}

The achievement of Bose-Einstein Condensation (BEC) in dilute gases has
been quickly followed by the demonstration of atomic outcouplers \cite{Mewes:1997,Hagley:1999,Bloch:1999,Cennini:2003},
allowing atoms to be extracted coherently, all in the same wavefunction, forming a so-called \textit{atom laser}.
Continuous outcoupling \cite{Bloch:1999,Cennini:2003} leads to quasicontinuous atom lasers, and among them gravity compensated guided atom lasers (GAL) \cite{nous:GAL,Couvert:2008,Kleine:2010} are promising tools to study quantum transmission through obstacles \cite{Carusotto:2001,Leboeuf:2001,Carusotto:2002,Paul:2005_1,Paul:2009PRA,Gattobigio:2009}, as they provide a large and constant de~Broglie wavelength along the propagation. Tuning the atom laser flux should allow one to investigate both linear and nonlinear phenomena,
 and even reach strongly correlated ones~\cite{Carusotto:2001}. However quasicontinuous emission imposes to operate in the so-called \emph{weak coupling regime} (see e.g.~\cite{Debs:2010} and references therein) such that corresponding flux limitations may restrict the envisaged applications. Thus,
this work aims to characterize in detail the flux properties of the horizontally, radiofrequency (rf) outcoupled, GAL demonstrated in \cite{nous:GAL}, in order to identify the limits of the weak coupling regime.

The weak coupling limits are intimately linked to the irreversibility of the coupling process; the coupling strength has to be weak enough for the extracted atoms not to be coupled back into the trapped BEC \cite{Moy:1999,Jack:1999,gerbier:2001}. In particular, the emergence of bound states as the coupling strength is increased constitutes a major limitation \cite{Hope:2000,Jeffers:2000} and can even result in the shut-down of the atom laser \cite{Robins:2005,Robins:2006}. In this respect, the crucial role played by the underlying force in the coupling process (arising either from the gravity acceleration in the case of free space rf outcoupling or from the two-photon momentum kick in the case of Raman outcoupling) has been stressed \cite{Debs:2010}. This force pushes out the atoms and prevents them to be recaptured before they have left the coupling zone.

For the horizontally GAL studied here, gravity is cancelled, and the extracting force is only provided by the repulsive atomic interactions with the remaining trapped BEC. This has important consequences on the flux properties. First, the outcoupling spectrum, i.e.~the dependence of the output flux versus the rf frequency, is profoundly modified. From a calculation based on the Fermi golden rule, we indeed show that it exhibits a double structure, with a large pedestal and a very narrow peak around its maximum. This is in stark contrast to rf-outcoupled free falling atom lasers~\cite{Bloch:1999}. In practice, it results in a strong sensitivity to residual magnetic fluctuations, which should be suppressed as much as possible. Second, the horizontally GAL is subjected to severe flux limitations, as expected from the weakness of the extracting force. In particular, we explicit here the link between such limitations, arising from the onset of bound states in the process as recognized in other schemes \cite{Robins:2005,Robins:2006,Debs:2010}, and the so-called ``quantum pressure'' associated with the confinement of the atom laser wave function around the outcoupling zone. Altogether, the maximal flux achievable is typically one order of magnitude less than for the gravity driven atom lasers. However these limitations do not impair future applications for studying quantum transport phenomena, as we finally discuss.

Both aspects (the outcoupling spectrum and the weak coupling limits) are investigated using a simplified quasi-1D model, whose predictions are compared, with an overall fair
agreement, to the experiments. The paper is organized as follows. We present in section~\ref{modeling} the scheme of the GAL demonstrated in~\cite{nous:GAL} and the quasi-1D model we use. Then we study in section~\ref{par.spectrum} the outcoupling spectrum and detail in section~\ref{par.weakcoupling} the weak coupling limits. Finally we conclude and discuss some prospects in section~\ref{par.discussion}.

\section{\label{modeling} The rf guided atom laser: principle and modelling}

\subsection{\label{par.geometry} Geometry and outcoupling scheme}

\begin{figure}[t]
\centering
\includegraphics[width=9cm]{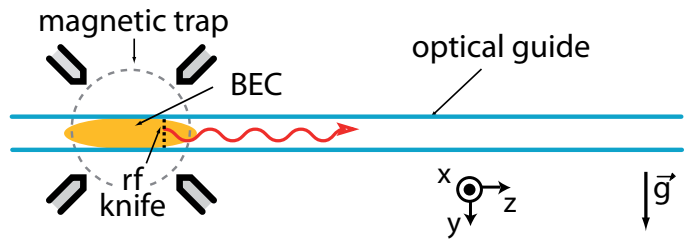}
\caption{Generation of an atom laser by rf outcoupling from a BEC created in an hybrid opto-magnetic trap. The optical guide is horizontal such that the atoms do not undergo any acceleration after leaving the BEC zone. Note that an atom laser is emitted on both side. For clarity, only one is shown here.}
\label{fig.manip}
\end{figure}

As depicted in figure~\ref{fig.manip}, a BEC containing $N_{0}$ $^{87}$Rb atoms is directly created inside an optomagnetic trap in the $|F, m_{F}\rangle=|1, -1 \rangle$ magnetic state. This cigar-shaped trap is the combination of an horizontal optical guide, which generates
a strong cylindrical transverse confinement (with a radial frequency $\omega_{\perp}$), and a quadrupole magnetic field, which provides a weak longitudinal confinement (of frequency $\omega_{z}\ll \omega_{\perp}$) along the $z$ axis. The chemical potential of the BEC verifies $\mu_{\BEC}\gg \hbar\omega_{\perp}$, such that the BEC is well in the 3D Thomas-Fermi (TF) regime~\cite{Dalfovo:1999}. Its density $n_{\rm{3D}}(\rp,z)$ ($\rp$ being the radial coordinate) has then an inverted parabola shape of radii $R_{\perp,z} =(2\mu_{\BEC}/m\omega_{\perp,z}^2)^{1/2}$, with $m$ the atomic mass.

The atom laser is generated using an rf magnetic field to coherently spin-flip a small fraction of the atoms from the source BEC
to the $|1,0\rangle$ state which is insensitive (at first order) to the magnetic field. While the outcoupled atoms are still radially confined by the optical guide, they can move along the longitudinal axis $z$. More precisely, they feel the sum potential $V_{\rm{g}} (\rp,z) + g n_{\rm{3D}}(\rp,z)$. Here $ V_{\rm{g}}$ takes into account the optical guide confinement together with the second order Zeeman quadratic effect, and the second term corresponds to the mean-field interaction with the remaining trapped BEC.
Here $g=4\pi a \hbar^2/m$ is the collisional parameter and $a$ the scattering length.

This interaction provides the necessary force to extract the atoms from the coupling zone. The atoms accelerate down the ``bump-shaped" mean-field potential until they leave the BEC region and then propagate at constant velocity $v$, i.e.~constant de Broglie wavelength $\lambda_{\rm{dB}}=h/mv$~\cite{nous:GAL}.

\subsection{\label{par.1Dmodel}Hypothesis and quasi-1D model}

In the following, we use a simplified model to describe the GAL. It relies on the assumptions listed below:
\begin{enumerate}
\item The influence of the $|1,1\rangle$ magnetic state is negligible and we only consider a two-level system (see section~\ref{par.expcouplage}).
\item  The BEC wave function evolves adiabatically during the outcoupling process. Apart from section~\ref{par.weakcouplingexp}, we also neglect the depletion of the BEC.
\item The collisional parameter $g$ is independent of
the magnetic substate, which is approximatively the case for $^{87}$Rb \cite{Burke:1997}.
\item  The intra-laser interactions are negligible in the outcoupling process. Note that this does not prevent the interactions to play a significant role along the propagation in the guide (see section~\ref{par.discussion}).
\item The atom laser does not experience any residual force along the guide axis $z$: $ V_{\rm{g}}(\rp,z)=m\omega_{\perp}^{2}\rp^{2}/2 $. This can be achieved by using the weak longitudinal optical guide confinement to compensate precisely the second-order Zeeman effect~\cite{nous:GAL}.
\end{enumerate}

Besides, we adapt the usual quasi-1D approximation (see e.g.~\cite{Jackson:1998}) to our configuration where two magnetic substates, the atom laser and the BEC, are involved. If one consider only one single state, the quasi-1D approximation supposes that the transverse wavefunction adjusts adiabatically along the
guide axis: it sticks to the ground state of the local transverse potential, which depends on the \textit{total} mean-field interaction, i.e. including both the inter- and intra-states contributions. In our case, the hypothesis of $g$ being independent of the magnetic substates is of particular importance: it implies that the atom laser and the BEC experience the same transverse potential. Their transverse wave function should thus be identical, and the same as if all atoms were in the same state. Consistently, we assume a perfect transverse mode matching and write the total wave function $\psit (\bi{r},t)$ in the separate form
\begin{eqnarray}
\psit= \left[
\begin{array}{c}
\psib(\bi{r},t) \\
\psil(\bi{r},t)
\end{array}
\right]= \left[
\begin{array}{c}
\phib(z,t) \\
\phil(z,t)
\end{array}\right]\;\psip(\rp, z) \; ,\label{Eq.quasi1D}
\end{eqnarray}
with the normalization condition $\int |\psit|^2 d \rp dz =1$. Here the subscript $\BECA$ and $\rm{L}$ refers to the BEC and the atom laser respectively. Neglecting the intra-laser interactions, $\psi_\perp$ evolves continuously from the transverse inverted parabola shape of the BEC to the gaussian fundamental mode of the guide.

\begin{figure}[b]
\centering
\includegraphics[width=9cm]{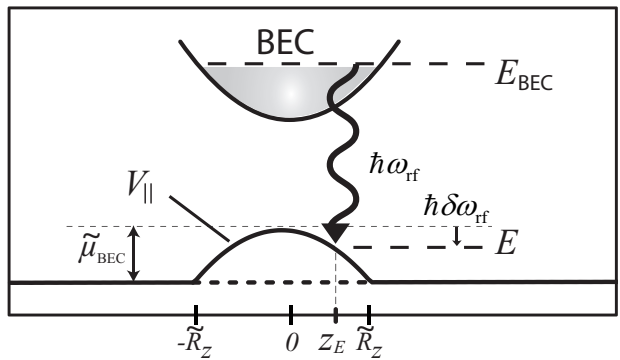} \caption{Effective 1D energy diagram for the outcoupling process. $V_\parallel (z)$ [equation~(\ref{Eq.Vparallel})] is the potential seen by the atom laser state. $\tilde{\mu}_{\BEC}=\mu_{\BEC}-\hbar\omega_{\perp}$ is the corrected chemical potential and $\tilde{R}_{z}$ the BEC edge. $\delta\omega_{\rf}$ is the frequency detuning from a coupling at the center of the condensate and $z_{E}$ corresponds to the classical turning point given by equation~(\ref{Eq.zE}) (only the $z>0$ resonance is shown). } \label{fig.resonance}
\end{figure}

As detailed in \ref{appendix_1D}, the outcoupling process is then reduced to a one-dimensional problem. In particular, the effective 1D mean-field potential felt by the atom laser follows the BEC profile (see figure~\ref{fig.resonance}) and reads
\begin{equation}
V_{\parallel} (z) = \hbar\omega_{\perp} \,+\,\tilde{\mu}_{\BEC}\Big(1- \frac{z^2}{\tilde{R}_{z}^{2}}\Big) \, \Theta \Big(1-\frac{ |z|}{\tilde{R}_{z}} \Big)
\label{Eq.Vparallel}
\end{equation}
the origin of the $z$-axis being at the center of the BEC. Here $\Theta$ is the Heaviside step function, $\tilde{\mu}_{\BEC}=\mu_{\BEC}-\hbar\omega_{\perp}$ is the chemical potential corrected from the transverse fundamental mode energy and $\tilde{R}_{z}=(2\tmu_{\BEC}/m\omega_{z}^2) ^{1/2}$. The latter corresponds precisely to the BEC edge in the quasi-1D approximation. In the following, we omit the constant in~(\ref{Eq.Vparallel}) and use the implicit energy shift $E \rightarrow E-\hbar\omega_{\perp}$.

\section{\label{par.spectrum}Outcoupling spectrum: flux \textit{vs} rf frequency }

We study now the dependence of the flux \textit{versus} the rf frequency, i.e.~the outcoupling spectrum. In particular, we show that it exhibits a very peaked shape. Note that we assume throughout this section that the atom laser operates within the weak coupling regime, which is defined below. The associated conditions will be detailed in the next section.

\subsection{Theoretical framework}

As sketched in figure~\ref{fig.resonance}, the rf outcoupling process can be seen as the coupling of an initial state (BEC) to the continuum formed by the \textit{stationary} states in the potential $V_{\parallel}(z)$. We consider here the weak coupling regime, which is defined by the regime where the coupling process is irreversible: the extracted atoms leave the coupling zone irretrievably, without any coherence with the remaining trapped BEC~\cite{Jack:1999,Moy:1999,gerbier:2001}. It is also commonly referred as the Born-Markov regime~\cite{Gardiner}. In this situation, the emission is quasicontinuous and the BEC state decays exponentially, the outcoupling rate $\Gamma$ (or equivalently the flux $\mc{F}=N_{0} \Gamma$) being given by the Fermi golden rule \cite{gerbier:2001}
\begin{equation}\label{Eq.GoldenRule}
\Gamma(E) = \frac{\pi\hbar\Omega_{\rf}^{2}}{2} \eta(E)
 \left| \int \phib (z)^{\ast}\, \phi_{\rm{L},E} (z)\; dz \right|^{2} .
\end{equation}
In this expression, $\eta(E)$ is
the 1D density of states and $\Omega_{\rf}$ quantifies the rf coupling strength (see \ref{appendix_1D}). $\phib$ corresponds to the longitudinal BEC wave function normalized to unity. It is well approximated by $|\phib(z)|\sim(15/16R_z)^{1/2}(1-z^{2}/R_z^{2} )\Theta(1-z/\tilde{R}_{z})$ except around the BEC edge ($z=\tilde{R}_{z}$), where it falls down to zero. Lastly, $\phi_{l,E} $ is the atom laser stationary wave function, also normalized to unity, whose energy $E$ is selected by the rf frequency $\omega_{\rf}/2\pi$ via the resonance condition
 \begin{equation}
E = E_{\rm{BEC}} - \hbar \omega_{\rf}\; .   \label{Eq.resonance}
\end{equation}
The calculation of $\Gamma(E)$ according to~(\ref{Eq.GoldenRule}) reduces then to the determination of $\phi_{\rm{L},E}$ (see \ref{App.ALwave}). However, in most cases, a semiclassical approach is
sufficient to obtain an accurate approximation~\cite{gerbier:2001}.

\subsection{Semiclassical approach}

Following the so-called Franck-Condon principle, we consider here
that the outcoupling is spatially localized at
the points defined by energy and momentum conservation during the
coupling, i.e., the classical turning points~\cite{Band:1999}. As both
the initial kinetic energy of the condensed atoms and the momentum
transfered by the rf field are negligible, the turning points are defined by $V(z_{E})=E$. It gives the two symmetrical positions
\begin{equation}\label{Eq.zE}
z_{E} = \pm \sqrt{\frac{2\hbar \, \omegarf}{m\omega_{z}^{2}}} \,,
\end{equation}
where $\omegarf = \omega_{\rf} - (E_{\rm{BEC}} - \tilde{\mu}_{\BEC})/\hbar$ is the frequency detuning from a coupling at the center of the
condensate (see figure~\ref{fig.resonance}). Formally, this approach is equivalent to approximate the atom laser wave
function $\phi_{\rm{L},E}(z)$ by a Dirac distribution around $z_E$, such that equation~(\ref{Eq.GoldenRule}) transforms into \cite{gerbier:thesis}
\begin{equation} \label{Eq.FranckCondon}
\Gamma_{\rm{sc}}(E) = \frac{\pi\hbar\Omega_{\rf}^{2}}{2}
\,  \int
dz\, |\phib(z)|^2 \, \delta\left(V_{\parallel}(z)-E \right).
\end{equation}
Considering the emission at one side, it finally reduces to the same expression as in~\cite{gerbier:2001}
\begin{equation} \label{Eq.FluxSC}
\Gamma_{\rm{sc}}(E)  = \frac{\pi\hbar\Omega^2_{\rf}}{2m\omega_{z}^{2} z_{E}}
 |\phib( z_{E})|^{2} \;,
\end{equation}
the gravity acceleration being replaced here
by the local acceleration $g_{\rm{eff}}=\omega_{z}^{2}z_{E} $ of the potential $V_{\parallel}$ at the position $z_{E}$. Consistently it vanishes for a coupling at the BEC edge ($z_{E}=\tilde{R}_{z}$ or equivalently $\omegarf=\tilde{\mu}_{\BEC}/\hbar$), the corrected chemical potential being the outcoupling spectrum width in the semiclassical approach (see figure~\ref{fig.resonance}).

This approach is well justified when one outcouples at the BEC side. There the acceleration is large enough and the first lobe of the atom laser wave function (which can be approximated by an Airy function) is narrow compared to the size of the BEC (see \ref{App.ALwave}). However it fails near the top of the BEC ($g_{\rm{eff}} \rightarrow 0$), where the coupling rate should precisely reach its maximum. This region is thus of particular interest and calls for a more complete description.

\subsection{\label{GoldenRule} Outcoupling spectrum}

\begin{figure}[t]
\centering
\includegraphics[height=7.5cm]{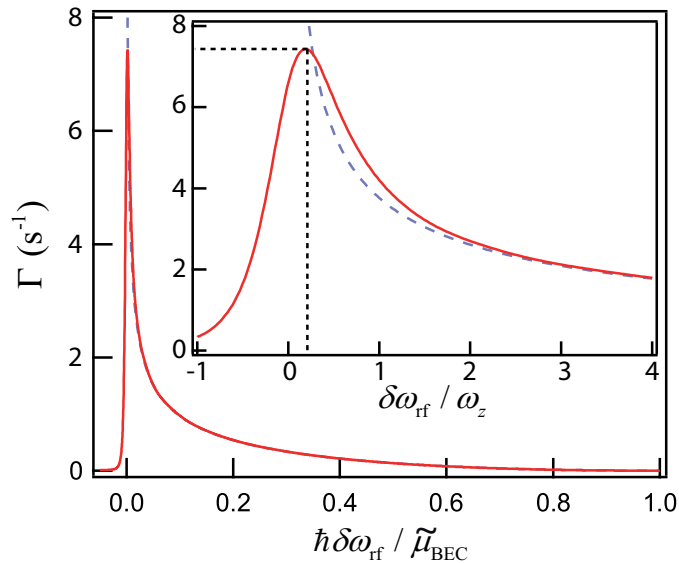}
\caption{Outcoupling rate versus the normalized detuning $\hbar\omegarf/\tmu_{\BEC}$. Straight (red) line: outcoupling rate $\Gamma$ given by the Fermi golden rule [equation~(\ref{Eq.GoldenRule})]. Dashed (blue) line: semiclassical
approximation $\Gammasc$ [equation~(\ref{Eq.FluxSC})]. Inset: zoom on
the vicinity of the zero detuning (the normalization is changed to $\omega_z$). The parameters ($\tmu_{\BEC} = 2.5$~kHz, $\omega_z/2\pi=$27~Hz and $\Omega_{\rf}/2\pi=15$~Hz) are those of section~\ref{par.expcouplage}.}
\label{fig.couplingrate}
\end{figure}

The determination of the stationary atom laser wave function $\phi_{\rm{L},E}$ is detailed in the \ref{App.ALwave}. Then, using equation~(\ref{Eq.GoldenRule}), we calculate the outcoupling rate. The corresponding spectrum $\Gamma(\omegarf)$ is plotted in figure~\ref{fig.couplingrate}.

As expected the outcoupling spectrum extends over a typical width $\tilde{\mu}_{\BEC}/\hbar$ and agrees well with the semiclassical expression $\Gammasc$ when coupling at the BEC side ($\omegarf \gg \omega_z$). \textit{De facto}, its validity range is very broad and a significative deviation only appears as $\omegarf\sim \omega_z$. Such behaviour can be qualitatively understood noticing that the classical turning point $z_E$ approaches here
 the characteristic length $\sigma_z=(\hbar/m\omega_z)^{1/2}$ of the inverted harmonic potential [see equation~(\ref{Eq.zE})]. Thus the curvature dominates and leads to the saturation of the first lobe of the atom laser wave function $\phi_{\rm{L},E}$ (see \ref{App.ALwave}). We thus expect the outcoupling rate to reach a maximum around this position. Finally, it should fall down rapidly to zero for $\omegarf\lesssim0$, as it corresponds to the classically forbidden region. Such analysis is refined by a zoom on this region (see the inset of figure~\ref{fig.couplingrate}). In particular, it appears that the maximum value is reached around $\omegarf \sim \omega_z/4$ and
\begin{equation}
\max_{\omegarf}\big(\Gamma\big) \sim  \Gammasc \Big( \frac{\omegarf}{4}  \Big)=\frac{ \Omega_{\rf}^2} {\omega_z } \;\Pi_{\sigma_z} \;.\label{Eq.GammaMax}
\end{equation}
Here $\Pi_{\sigma_z} = \pi \sigma_z|\phi_{\BECA}(0)|^2 /\sqrt{2} $ corresponds to the proportion of atoms contained in the region of spatial width $\pi \sigma_z/\sqrt{2}$ \footnote{
Consistently, we recover the expression for a coupling to a continuum of finite width $\hbar \omega_z$, the Rabi oscillation frequency being $\Omega_{\rf} \sqrt{\Pi_{\sigma_z}}$ \cite{CohenAnglais:1992}.}.

Altogether, the outcoupling spectrum exhibits a double structure. The large one, of width given by the chemical potential $\tmu_{\BEC}$, corresponds to a coupling at the BEC side. The very narrow one, of width $\omega_z$, is associated to the sharp increased of $\Gamma$ around the BEC center. For a BEC in the 3D Thomas-Fermi regime ($\mu_{\BEC} \gg \hbar\omega_\perp \gg \hbar\omega_z$), it results in a very peaked outcoupling spectrum, in contrast with the common behaviour of free falling atom lasers (see e.g.~\cite{gerbier:2001}).

Such peaked shape, whose narrow width is typically around a few tens of Hertz, has important consequences for the use of such GAL. Indeed weak residual magnetic fluctuations are hardly avoidable in practice. First those will slightly broaden the spectrum such that the narrow peak should be hard to observe experimentally. Most importantly, those may result in strong density fluctuations when coupling near the BEC center. This indicates that an extreme attention should be paid when operating in this region. As we shall see in section~\ref{par.weakcoupling}, this statement is even more reinforced as this position corresponds to the most severe weak coupling conditions.

\subsection{\label{par.expcouplage}Experimental investigation}

We present now our experimental investigation on the outcoupling spectrum. The optical guide is created by a Nd:YAG laser ($\lambda$=1064~nm) focused on a waist of 30~$\mu$m, resulting in a radial trapping frequency $\omega_\perp/2\pi = 330$~Hz. The weak magnetic longitudinal confinement is set to $\omega_z/2\pi =
27$~Hz. The initial BEC contains $N_0=1.1\times 10^5$ atoms, corresponding to $\tmu_{\BEC}/h =
2.5$~kHz.

The bias field of our Ioffe-Pritchard magnetic trap is chosen around $B_0\sim 7$~G (see \cite{nous:RSI} for further details). For such bias, the Zeeman quadratic effect enables us to perform selectively the rf coupling to the atom laser state, without populating the $|1,1\rangle$ state~\cite{Desruelle:1999a}. We calibrate the coupling strength by using very short outcoupling pulses ($t_{\rf}\ll \hbar/\tmu_{\BEC}$). There the whole BEC oscillates precisely at the pulsation $\Omega_{\rm{rf}}$, which is determined within a few percent.

\begin{figure}[t]
\centering
\includegraphics[height=7.5cm]{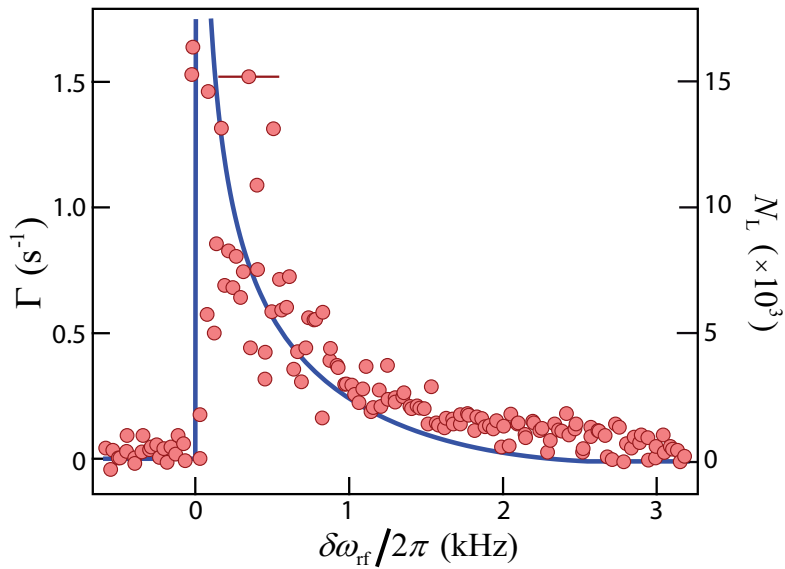}
\caption{Experimental outcoupling spectrum $\Gamma(\omegarf)$ for the coupling strength $\Omega_{\rf}/2\pi=15.1 \pm 0.5$~Hz.
Left vertical axis: outcoupled atom number $N_{\rm{L}}$ (for one side). Red circles: experimental results. Straight line: theoretical predictions given by equation~\ref{Eq.GoldenRule}. The total atom number is $N_0=1.1 \times 10^5$ and $\tmu_{\BEC}/h=$2.5~kHz. The horizontal error bars (200~Hz rms) corresponds to the uncertainty on the detuning caused by the shot-to-shot magnetic fluctuations. Note that the detuning has been corrected from the long term bias drift estimated around 0.4~mG/h (300~Hz/h).}
\label{fig.CourbeCouplage}
\end{figure}

In contrast, a quasicontinuous GAL is extracted from the BEC by applying a long and weak rf pulse. In the following, the coupling time is set to $t_{\rf}=50$~ms. It should allow us to resolve, in principle, the peaked shape of the outcoupling spectrum ($t_{\rf}\gg 1/\omega_z \gg\hbar/\tmu_{\BEC}$). An experimental spectrum is shown in figure~\ref{fig.CourbeCouplage} for a coupling strength $\Omega_{\rf}/2\pi=15.1$~Hz. The extracted atom number for one side, $N_{\rm{L}}$, is measured by fluorescence imaging and used to calculate the outcoupling rate according to $\Gamma=N_{\rm{L}}/ N_0\, t_{\rf}$, the depletion of the BEC being negligible for our parameters.

For a coupling at the BEC side, we find a fair agreement with the predictions (without any free parameters). For a coupling near the BEC edge, we observe a slight deviation that can be traced to the presence of residual thermal atoms. This broadens the spectrum, leading to a total width ($\sim$3~kHz) slightly above the ideal value given by the corrected chemical potential $\tmu_{\BEC}/h$.

Looking at the coupling near the zero detuning, the situation is balanced. On the one hand, we recover a peak-shaped structure associated with a sharp variation of the flux. It shows that the rapid magnetic fluctuations have been fairly suppressed. In particular, one can infer an upper bound around $70$~$\mu$G rms (50~Hz converted in frequency unit). This was achieved by using a magnetic shielding and actively stabilizing the current $I_0$ that creates the bias field ($\Delta I/I_0\sim10$~ppm rms).

On the other hand, the spectrum is ``scrambled'' in this region by the remaining shot-to-shot fluctuations. Those can be estimated around $250$~$\mu$mG rms (200~Hz) and prevent at the moment a detailed investigation of the narrow peaked structure. \textit{De facto}, the observed maximum outcoupling rate lies around $\Gamma\sim1.5$~s$^{-1}$, still far from the expected maximum (see figure~\ref{fig.couplingrate}). As we shall see in the next section, let us stress that this difference may very unlikely be attributed to saturation effects beyond the weak coupling regime.

\section{\label{par.weakcoupling}Flux limitations for the rf guided atom laser}

In this section, we explicit the weak coupling limits for the GAL. We show that they originates from the emergence of bound states in the coupling process and we especially emphasize their close link with the ``quantum pressure'' for rf coupling schemes.

\subsection{\label{par.conditions}Weak coupling conditions}

As already mentioned, the weak coupling regime is referred as the situation where the coupling strength is low enough for the dynamics to be irreversible, i.e.~well described by the Born Markov approximation. This imposes several conditions associated with different physical phenomena.

Building upon the modeling of the outcoupling process as the coupling to a continuum, one can first derive a necessary condition: the memory time $t_{\rm{m}}$ of the continuum has to be short enough for the extracted atoms not to be coupled back to the trapped state, i.e.~$\Gamma t_{\rm{m}}\ll 1$~\cite{Jack:1999,Moy:1999}. This finite time $t_{\rm{m}}$ being linked to the finite \textit{effective }width of the continuum, $t_{\rm{m}} \sim \hbar/\Delta$, the condition can be equivalently written in the canonical form $\Gamma \ll \Delta $ \cite{CohenAnglais:1992,gerbier:2001}. From the double structure of the outcoupling spectrum discussed in section~\ref{GoldenRule}, we expect this limit to evolve from $\Gamma\ll\omega_z$ for an outcoupling near the BEC center to $\Gamma\ll \tmu_{\BEC}/\hbar$ at the BEC side.

However, a more strict condition arises from the emergence of bound states in the rf dressed basis, as first mentioned by Stenholm \textit{et al.} \cite{Stenholm:1997,Paloviita:1997} in the context of laser induced molecular excitations. If the coupling strength is increased above a certain critical value, part of the atoms remains indeed trapped in a coherent superposition of the atom laser and BEC states, and do not escape the coupling region. Such phenomenon leads to a strong modification of the outcoupling dynamics \cite{Hope:2000,Jeffers:2000} and ultimately results in the shut-down of the atom laser, as observed experimentally \cite{Robins:2005,Robins:2006,Debs:2010}.

\begin{figure}[b]
\centering
\includegraphics[width=10cm]{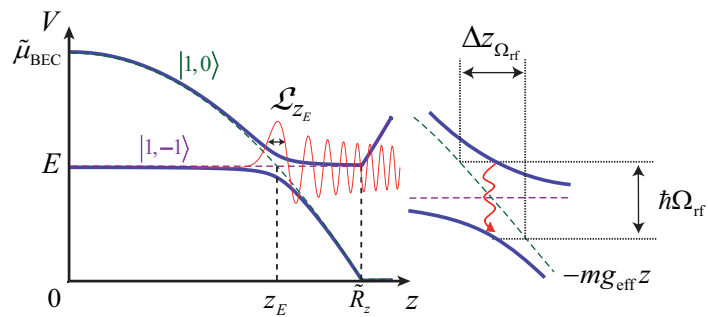}
\caption{Energy level crossing in the dressed state basis at given rf detuning $\omegarf$ and coupling strength $\Omega_{\rf}$. Dashed lines: bare potentials for the BEC ($|1,-1\rangle$) and atom laser ($|1,0\rangle$) states. Solid lines: dressed state potentials. The crossing position $z_E$ corresponds to the classical turning point and $g_{\rm{eff}}=\omega_z^2 z_E$ the local acceleration. The typical anti-crossing extension is given by $\Delta z_{\Omega_{\rf}}=\hbar\Omega_{\rf}/m\geff$. Red line: stationary atom laser wave function $\phi_{\rm{L},E}$. $\mathcal{L}_{z_E}$ is the typical extension of the first lobe (see text).}
\label{fig.dressedstate}
\end{figure}

Following the analysis of Debs \textit{et al.}~\cite{Debs:2010}, the critical coupling strength $\omegac$ coincides with the adiabatic following condition in the rf dressed potentials. These potentials are shown in figure~\ref{fig.dressedstate} for an outcoupling at the BEC side. In this case, the local acceleration $\geff=\omega_z^2 z_E$ is finite at the anti-crossing position [i.e.~the classical turning point $z_E$ given by equation~(\ref{Eq.zE})] and the results of Ref.~\cite{Zobay:2004} apply. From the transfer probability between the two dressed states \footnote{The transfer probablity $P\sim\rm{exp}-(\Omega_{\rf}/\omegac)^{3/2}$ of Ref.~\cite{Zobay:2004} can be seen as the extrapolation of the Landau-Zener formula for a particle starting at zero velocity at the crossing position.}, one can therefore infer the weak coupling limit \begin{equation}
\Omega_{\rf}^{\rm{c}}\sim \Big(\frac{2mg^2_{\rm{eff}}}{\pi^2\hbar}\Big)^{1/3}=\Big(\frac{2}{\pi}\Big)^{2/3} \Big(\omegarf\omega_z^2  \Big)^{1/3}\;\label{Eq.omegac}
\end{equation}
for $\omegarf\gg \omega_z$. Such expression enlightens particularly well the importance of having a large local acceleration, that pushes out the extracted atoms and prevents their recapture, to achieve a large critical coupling, i.e. a large flux \cite{Debs:2010}.

Interestingly, one finds that the anti-crossing extension (see inset of figure~\ref{fig.dressedstate}) corresponding to the critical coupling strength $\Delta z_{\omegac}=\hbar\omegac/m\geff$ coincides precisely to the typical Airy lobe size $\mathcal{L}_{z_E}$ of the stationary atom laser wavefunction $\phi_{\rm{L},E}$: $\Delta z_{\omegac}=(\hbar^2/2m^2g_{\rm{eff}})^{1/3}\equiv \mathcal{L}_{z_E}$.

This suggests an alternative picture to recover the limit given by equation~(\ref{Eq.omegac}). As a matter of fact, the dressed state basis diagonalizes the bare potentials and coupling at each position, i.e.~the complete Hamiltonian, except for the kinetic energy. Formally, it means that the kinetic energy term drives the coupling between the two dressed states. The decay from one state to the other is then governed by its relative importance compared to the coupling strength (see e.g.~\cite{Zobay:2004}). The adiabatic following condition can thus be reformulated as $E_k(z_E)\ll\hbar \Omega_{\rf}^{\rm{c}}$, where $E_k(z_E)$ is the kinetic energy at the crossing position.
In the case of a rf outcoupler, no momentum kick is imparted to the atoms and, in a semiclassical picture, they leave the outcoupling zone without initial velocity. However $E_k(z_E)\sim \hbar^2/2m \mathcal{L}_{z_E}^2$ remains finite and is given by the so-called quantum pressure associated to initial confinement. This leads directly to the criterion~(\ref{Eq.omegac}) (apart from a scaling prefactor $(2/\pi)^{2/3}$ close to unity).

For an outcoupling near the BEC center, the local acceleration vanishes and the validity of~(\ref{Eq.omegac}) breaks down. However, the above approach based on the quantum pressure can be directly extended in this region, the Airy lobe size being replaced by the harmonic oscillator size $\sigma_z$. The critical coupling writes then
\begin{equation}\Omega_{\rf}^{\rm{c}}\sim\frac{\omega_z}{(2\pi^2)^{1/3}} \;\label{Eq.limitcenter}\end{equation} for $\omegarf \lesssim \omega_z$ (for consistency, the scaling prefactor found above has been included). Note that it corresponds to the limit found in \cite{Hope:2000,Jeffers:2000}, considering non-interacting atoms coupled into free space.

Altogether, we bridge the gap between the two domains using the phenomenological expression
\begin{equation}
\Omega_{\rm{rf}}^{\rm{c}} \,\big(\omegarf\big)=\frac{\omega_z}{(2\pi^2)^{1/3}} \,\Big[ \;1\,+\, 64\,\Big(\frac{\omegarf}{\omega_z}\Big)^{2}\;\Big]^{1/6}\;\label{Eq.CouplageCritique}
\end{equation}
 over the whole coupling spectrum [see figure~\ref{fig.maximalcoupling} (a)]

\subsection{Critical outcoupling spectrum}

\begin{figure}[t]
\centering
\includegraphics[width=15cm]{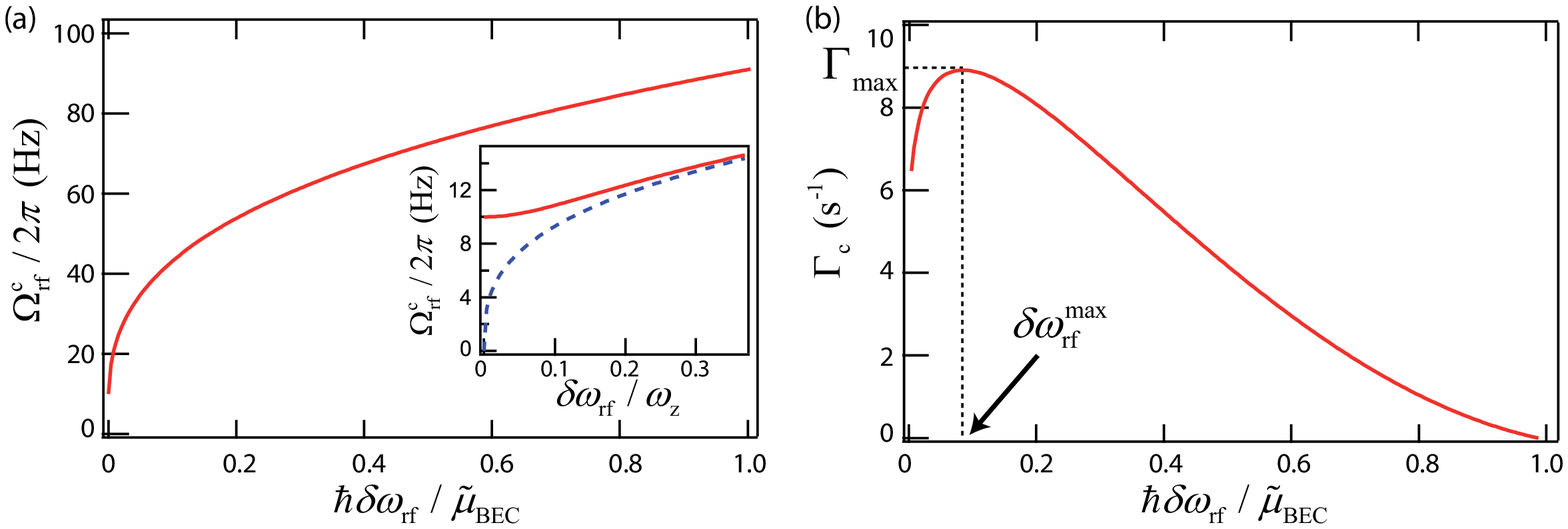}
\caption{(a) Evolution of the critical coupling strength $\omegac$ with the normalized outcoupling detuning $\hbar\omegarf/\tmu_{\BEC}$. Straight and dashed lines: $\omegac$ given by equations.~(\ref{Eq.CouplageCritique}) and (\ref{Eq.omegac}) respectively. Inset: zoom around the zero detuning, with a normalization changed to $\omega_z$. (b) Critical outcoupling spectrum $\Gammac(\omegarf)=\Gammasc(\omegac,\omegarf)$ for the detuning range $\omegarf \in[\omega_z/4\,,\,\tmu_{\BEC}/\hbar]$. $\Gamma_{\rm{max}} $ is the maximal coupling rate and $\delta\omega_{\rf}^{max}$ the associated detuning. The parameters are the ones of section~\ref{par.expcouplage}: $\omega_z/2\pi=27$~Hz and $\tmu_{\BEC}/h$=2.5~kHz. }
\label{fig.maximalcoupling}
\end{figure}

From the limit given by equation~(\ref{Eq.CouplageCritique}), we deduce the critical outcoupling rate $\Gammac  = \Gammasc\big(\omegac,\omegarf\big)$ for each detuning. Note that we use here the semiclassical expression~(\ref{Eq.FluxSC}) and restrain the detuning range to $\omegarf \in[\omega_z/4\,,\,\tmu_{\BEC}/\hbar]$ where this approximation is accurate. The resulting critical outcoupling spectrum is plotted on figure~\ref{fig.maximalcoupling} (b). Here the very peaked structure has disappeared: the sharp decrease of $\Gammasc(\omegarf)$ is indeed partly compensated by the concomitant increase of $\omegac (\omegarf)$. In particular the evolution of $\Gammac$ is rather smooth around its maximum $\Gamma_{\rm{max}}$,
i.e.~the overall maximum outcoupling rate achievable within the weak coupling regime. Such maximum reads
\begin{equation}
\Gamma_{\rm{max}}\equiv  \max_{\omegarf} \Big\{\Gammac (\omegarf)\Big\} \sim 0.22 \frac{\omega_z }{(\tmu_{\BEC}/\hbar\omega_z)^{1/3}} \;,
\label{Eq.FLuxMax}
\end{equation}
the optimal outcoupling position $\delta\omega_{\rf}^{\rm{max}}\sim 0.08\,\tmu_{\BEC}/\hbar $ being shifted towards the BEC edge \footnote{Note that $\Gamma(\omegarf)$ being less sensitive to residual fluctuations in this region (see section~\ref{GoldenRule}), the corresponding detuning $\delta\omega_{\rf}^{\rm{max}}$ is also optimal from a practical point of view.}.

The above upper bound~(\ref{Eq.FLuxMax}) is of particular importance as it allows us to explicitly define the operating range of the atom laser. It imposes a strict limitation on the outcoupling rate, having in particular $\Gamma_{\rm{max}}\ll \omega_z$. Considering the experimental parameters of the previous section, one finds $\Gamma_{\rm{max}}\sim 8$~s$^{-1}$, i.e.~a maximum flux $\mc{F}_{\rm{max}}=N_0 \, \Gamma_{\rm{max}}\sim 10^6$~at.s$^{-1}$. This is typically one or two orders of magnitude less than with ``common'' schemes (see e.g.~\cite{Debs:2010}). As already said, such low flux operation was expected for the GAL. It can be traced to the very weak extracting force (only due to the interaction with the BEC) experienced by the atoms.

Finally, we can estimate the critical
coupling strength for a detuning $\omegarf\sim\omega_z/4$ corresponding to the maximum outcoupling rate in the weak coupling regime (see figure~\ref{fig.couplingrate}). As shown in the inset of figure~\ref{fig.maximalcoupling}, we find $\omegac\sim$13~Hz around this position, leading to $\Gammac \sim 6$~s$^{-1}$. The coupling strength used to calculate the outcoupling spectrum presented in figure~\ref{fig.couplingrate} being slightly above such critical value ($\Omega_{\rf}=15.1$~Hz), effects beyond the weak coupling regime should \textit{a priori} be taken into account in the predictions. However those may unlikely explain the experimental observations shown in figure~\ref{fig.CourbeCouplage}, the apparent saturation being most certainly due the residual shot-to-shot magnetic fluctuations.

\subsection{\label{par.weakcouplingexp} Experimental investigation}

\begin{figure}[t]
\centering
\includegraphics[width=9cm]{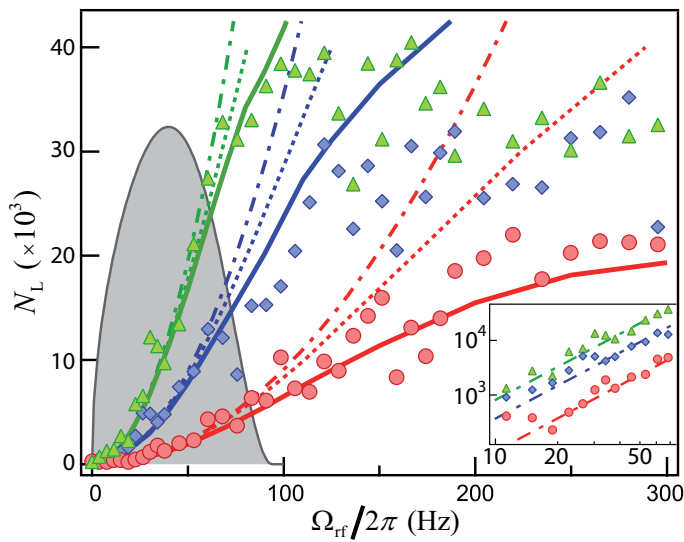}
\caption{Experimental investigation of the weak coupling limits: the extracted atom number for one side $N_{\rm{L}}$ \textit{vs} the coupling strength $\Omega_{\rf}$ for different detunings. $\triangle$ (green): $\omegarf/2\pi= 0.6 \,\pm 0.1$~kHz. $\diamond$ (blue): $\omegarf/2\pi= 1.3\, \pm 0.1 $~kHz. $\circ$ (red): $\omegarf/2\pi= 2.3 \,\pm0.1 $~kHz. The outcoupling time is $t_{\rf}=20$~ms. Dash-dotted lines: semiclassical prediction $N_{\rm{L}}=\Gammasc(\omegarf)\,t_{\rf}$. Dotted lines : the same with the depletion of the BEC being taken into account (see text). Thick lines: numerical resolution of the coupled quasi-1D GP equations. The grey area corresponds to the weak coupling regime defined by equation~(\ref{Eq.CouplageCritique}). Its border (thin line) is given by the parametric curve $[N_{\rm{L}}^{\rm{c}}(\omegarf),\omegac(\omegarf)]$, $N_{\rm{L}}^{\rm{c}}=\Gammac (\omegarf)t_{\rf}$ being the critical outcoupled atom number. The inset is a zoom, in log-log scale, on low rf coupling strength.} \label{fig.weakcoupling}
\end{figure}

Here we investigate experimentally the flux limitations by monitoring the extracted atom number $N_{\rm{L}}$ (still for one side) as the coupling strength $\Omega_{\rf}$ is progressively increased. The trap frequencies are the same as in section~\ref{par.expcouplage} and the initial atom number is now $N_0=2\times 10^5$~atoms, leading to $\tmu_{\BEC}/h=3.3$~ kHz. The outcoupling time is decreased to $t_{\rf}=20$~ms in order to keep the BEC depletion small enough for large outcoupling rate. The results are shown in figure~\ref{fig.weakcoupling} for three different detunings ($\omegarf/2\pi=0.6$, 1.3 and 2.3~kHz). Note that they correspond to outcoupling positions at the BEC side, where the extracted atom number is less sensitive to the magnetic shot-to-shot fluctuations.

As expected $N_{\rm{L}}$ agrees with the semiclassical predictions \footnote{Using either the Fermi golden rule~(\ref{Eq.GoldenRule}) or the semiclassical expression~(\ref{Eq.FluxSC}) gives very negligible differences for our parameters.} for weak coupling strengths (see inset of figure~\ref{fig.weakcoupling}). However, a significative deviation appears when $\Omega_{\rf}$ is increased and $N_{\rm{L}}$ starts to saturate. If a part of the deviation can be imparted to the depletion of the BEC, it cannot fully explain the observations. This is confirmed by comparing the experimental results with the solution \footnote{An analytical solution can be derived as detailed in~\cite{Bernard:thesis}.} of $\rm{d}N_{\rm{L}}/\rm{d}t= \Gammasc(N_{\BECA}) N_{\BECA}$, $\Gammasc$ being still given by equation~(\ref{Eq.FluxSC}) and $N_{\BECA}=N_0-2N_{\rm{L}}$ being the BEC atom number. In contrast, the numerical resolution of the coupled quasi-1D Gross-Pitaevskii (GP) equations [equations.~(\ref{Eq.GPB}) and ~(\ref{Eq.GPL})] fits well the experimental data, especially for the detuning $\delta\nu_{\rf}=2.25$~kHz (red curve). The saturation at large $N_{\rm{L}}$ for the detunings $\delta\nu_{\rf}=0.6$ and 1.29~kHz (green and blue curves) will be discussed below.

The disagreement between the experiments and our calculations based on the Fermi golden rule can be interpreted as a signature of the breakdown of the weak coupling regime. As a matter of fact, the onset of such deviation is well rendered, without any adjustable parameters, by the limit defined by equation~(\ref{Eq.CouplageCritique}). Thus, these observations comfort the analysis presented in section~\ref{par.conditions}: as previously stressed for other schemes \cite{Robins:2005,Robins:2006,Debs:2010}, the predominant flux limitation likely originates from the presence of bound states in the rf dressed potentials. As already said, equation~(\ref{Eq.CouplageCritique}) finally allows us to properly define the quasicontinuous operating range of the guided atom laser, which is symbolized by the gray area in figure~\ref{fig.weakcoupling}. In particular, it gives a maximal $N_{\rm{L}}$ around $3\times 10^4$ atoms for the parameters considered in this section, corresponding to a maximal flux $\mathcal{F}_{\rm{max}}\sim 1.5 \times 10^6$ at.s$^{-1}$.

Before concluding, one should stress that these results support \textit{a posteriori} the approximations made to model the GAL and to calculate the outcoupling rate. In particular we found no signatures (on the flux) of any BEC excitations in the weak coupling regime defined by~(\ref{Eq.CouplageCritique}) (grey area in figure~\ref{fig.weakcoupling}). The same applies for the intra-laser interactions. For the latter, their influence is characterized by the dimensionless parameter $a\nl$, where $\nl$ is the atom laser linear density. In the guide, it can be estimated using $\nl\sim \mc{F}/v$. Considering the typical propagation velocity $v\sim(2\tmu_{\BEC}/m)^{1/2} \sim 5$~mm.s$^{-1}$ together with the maximal flux $\mathcal{F}_{\rm{max}}$, one finds $(a\nl)_{\rm{max}} \sim 1.5$, i.e.~slightly above the limit of the 1D mean-field regime defined by $a\nl \lesssim1$~(see \ref{appendix_1D}). Increasing the flux further, the intra-laser interactions will profoundly modifies
the atom laser transverse wavefunction and so the transverse dynamic. Then our quasi-1D model may fail and it could explain the differences observed between the numerical simulations and the experiments at large outcoupled atom number (green and blue curves in figure~\ref{fig.weakcoupling}).

\section{\label{par.discussion}Conclusion and discussion}

We characterized in this paper the flux properties of the guided atom laser. In particular we showed that 	a quasicontinuous emission entails severe flux limitations ($\mathcal{F}\lesssim 10^6$ at.s$^{-1}$) compared to other schemes. However, it does not \textit{a priori} alter the very good promises for studying quantum transport phenomena, both in the linear and nonlinear regime. The inherent weak flux is indeed balanced by the very slow velocity propagation (and thus large de Broglie wavelength) offered by the rf outcoupling. It results in not so dilute atomic beams, typically within the 1D mean-field regime ($a\nl\leq 1$) as discussed above. In addition to single particle quantum effects (for instance quantum tunnelling or Anderson localization) a rich variety of nonlinear phenomena can be observed, as for instance the atomic blockade or a soliton generation past a finite lattice~\cite{Carusotto:2002}.

As a matter of fact, nonlinear effects may show up for very dilute atomic beams, as well illustrated by the transmission through a 1D disorder slab (see e.g.~\cite{Paul:2009PRA}). There the interaction-induced localization-delocalization crossover should be observed for the critical sound velocity $c=(2\hbar\omega_\perp a \nl/m)^{1/2}\sim v/7$, i.e.~deep in the supersonic regime. Taking $v=2$~mm.s$^{-1}$ as in \cite{Billy:2008}, it corresponds to an atomic density $a \nl \sim 0.05$ (in dimensionless unit), well within the limits given in this paper.

In future, the scheme described here could be extended to other configuration. Along these lines, one may implement Raman outcoupling techniques or head towards ``all optical'' techniques, similar to the one described in~\cite{Couvert:2008} but using Bragg transitions as an output coupler~\cite{Kozuma:1999}. Besides, tuning the scattering length through Feshbach resonances (using a proper atomic species) would certainly offer new possibilities (see e.g.~\cite{Duan:2010}).

To conclude, let us stress that the complete characterization of the rf outcoupled guided atom laser requires a detailed study of its longitudinal coherence, i.e.~its energy linewidth.
This will be the subject of a future work.


\ack

We thank J.-F. Riou for fruitful discussions. This research was supported by CNRS, Direction G\'en\'erale de l'Armement,
ANR-08-blan-0016-01, IXSEA, EuroQuasar program of
the EU. The Laboratoire
Charles Fabry de l'Institut d'Optique is a member of IFRAF.


\appendix

\section{\label{appendix_1D} Quasi-1D model}

We detail here the quasi-1D model for the guided atom laser. First we derive the quasi-1D coupled GP equations for the longitudinal wave functions $\phibA(z,t)$ and $\phil(z,t)$. Second we present the approximated form resulting from the assumptions listed in section~\ref{par.1Dmodel}.

\subsection{Quasi-1D coupled Gross-Pitaevskii equations}

The rf magnetic field $B_{\rf} \cos \omega_{\rf} t$ (along a direction perpendicular to the biais field $B_0$) couples the BEC and atom laser states.
In the rotating wave approximation, it results into the ``mean-field'' Hamiltonian
\begin{eqnarray}
\mathcal{H}= \left[
\begin{array}{cc}
 \frac{-\hbar^2}{2m}\triangle\!-\!\hbar\omegarf\!+ \!V_{\BECA}\! +\! \frac{g}{2}|\psit|^2& -\frac{\hbar\Omega_{\rf}}{2}\\
 -\frac{\hbar\Omega_{\rf}}{2} &\!\! \!\! \!\!\!\! \! \!\!\frac{-\hbar^2}{2m}\triangle+V_{\rm{g}}+ \frac{g}{2}|\psit|^2
\end{array}\!\right]\!\!,
\end{eqnarray}
where $\Omega_{\rf}=\mu_{\rm{B}}B_{\rf}/2\sqrt{2}$ is the coupling strength (taking the Land\'e factor $g_F=-1/2$). $V_{\BECA}=V_{\rm{g}}+V_{\parallel,\BECA}$ is the sum of the transverse potential $V_{\rm{g}}= \frac{1}{2}m\omega_\perp^2 \rp^2$ and the longitudinal confinement $V_{\parallel,\BECA}=\frac{1}{2}m\omega_z^2 z^2$. As discussed in section~\ref{par.1Dmodel}, we write the total wavefunction in the separate form
\begin{eqnarray}
\psit= \left[
\begin{array}{c}
 \phibA (z,t) \\
\phil(z,t)
\end{array}\right]\;\psip(\rp, \ntot) \; .\label{Eq.quasi1D}
\end{eqnarray}
with the normalization condition $\int |\psi_\perp|^2 d\rp  =1$. The global normalization is here $\int |\psit|^2 d\rp dz  =N_0$ such that $|\phibA|^2=\nbA$ and $|\phil |^2=\nl$ correspond respectively to the BEC and atom laser linear density. $\ntot=\nbA+\nl$ is the total linear density. Following the procedure of reference~\cite{Jackson:1998} (i.e.~minimizing the action associated to the functional energy $\langle \psit |\mathcal{H}  | \psit \rangle$), one obtains first the transverse stationary GP equation
\begin{equation}
\Big(- \frac{\hbar^2}{2m}\triangle_\perp  + V_{\rm{g}}  + g \ntot |\psip|^2\Big)\; \psip = \mu_{\rm{eff}}(\ntot)\; \psip\; , \label{transverse}
\end{equation}
 where $\mu_{\rm{eff}}$ is the effective local chemical potential. From equation~(\ref{transverse}), $\psi_\perp$ depends implicitly on $z$ via the total linear density, as written in~(\ref{Eq.quasi1D}). This property results from the hypothesis on $g$ being independent of the magnetic substates.

In the so-called ``1D mean-field'' regime ($a \ntot \ll 1$), the weak interactions do not perturb $\psip$, which sticks to the fundamental gaussian mode and $\mueff\sim \hbar\omega_\perp(1+2a\ntot)$. In the opposite strong density regime (3D TF regime defined by $a \ntot \gg 1$), $\psi_\perp$ is well approximated by an inverted parabola and $\mueff\sim2\hbar\omega_\perp\sqrt{a\ntot}$. Altogether the phenomenological expression
 \begin{equation}
 \mueff=\hbar\omega_\perp\sqrt{1+4a\ntot}\label{mueff}
 \end{equation}
is very accurate for any interaction strength \cite{Gerbier:2004}. Using equation~(\ref{mueff}), one finally derives the quasi-1D coupled GP equations for the longitudinal dynamic:
 \begin{eqnarray}
 i\hbar \frac{\phibA}{\partial t}&=&\Big(\!\frac{-\hbar^2}{2m} \triangle   \! -\!\hbar\delta\omega_{\rf}  \!+\!V_{\parallel,\BECA} \!+\! \mu_{\rm{eff}}\!\Big)\phibA\! - \!  \frac{\hbar\Omega_{\rf}}{2}\phil   \label{Eq.GPB}   \\
 i\hbar \frac{\phil}{\partial t} &=&\Big( \frac{-\hbar^2}{2m}\triangle  + \mu_{\rm{eff}}\,\Big)\, \phil    -   \frac{\hbar\Omega_{\rf}}{2}\,\phibA \; . \label{Eq.GPL} \end{eqnarray}
 The numerical simulations presented in figure~\ref{fig.weakcoupling} are obtained from this set of equations.

\subsection{Approximated form}

The above set of equations can be simplified assuming i) the adiabatic following of the BEC wave function and ii) no intra-laser interactions ($a\nl=0$). In the TF regime, equation~(\ref{Eq.GPB}) reduces indeed to
\begin{equation}
\mu_{\BEC}=\frac{1}{2}m\omega_z^2 z^2 + \hbar\omega_\perp\sqrt{1+4a\nbA(z)} \,,
\end{equation}
and the BEC linear density writes
\begin{eqnarray}
\nbA(z)&=&  \frac{1}{4a} \Big[  \Big(\frac{\mu_{\BEC}}{\hbar \omega_\perp}\Big)^2\, \Big( 1-\frac{z^2}{R_z^2}  \Big)^2\nonumber   \\
&&\;\;\;\; \;\;-\;1  \;    \Big]    \Theta \Big(1-\frac{ |z|}{\tilde{R}_z} \Big) \, .\label{Eq.nBtrue}
\end{eqnarray}
The BEC edge is located at $\tilde{R}_z=(2\tilde{\mu}_{\BEC}/m\omega_z^2)^{1/2}\sim R_z -\hbar\omega_\perp/2\mu_{\BEC}$, with $\tilde{\mu}_{\BEC}=\mu_{\BEC}-\hbar\omega_\perp$. Except around this position (where the density is extremely weak), it is very well approximated by $\nbA(z)\sim 15 N_0/16 R_z  (1-z^2/R_z^2 )^2$ ($z \ll \tilde{R_z}$), leading to the usual expression $\mu_{\BEC}=(15aN_0 \hbar^2\omega_\perp^2\omega_z)^{2/5}m^{1/5}/2$. Here the BEC depletion has been neglected but the time dependence can be included using $N_{\BECA}(t)=N_0-2N_{\rm{L}}(t)$ (where $N_{\rm{L}}$ is the atom number for one side) instead of $N_0$ in the preceding expressions.

From equation~(\ref{mueff}), the effective chemical potential reads finally
\begin{equation}
\mueff(z)= \hbar\omega_\perp +\tilde{\mu}_{\BEC}\Big(1- \frac{z^2}{\tilde{R}_z^2}\Big) \, \Theta \Big(1-\frac{ |z|}{\tilde{R}_z} \Big)\, .
\end{equation}
It acts as an effective 1D potential for the atom laser ($V_\parallel\equiv\mueff$), whose dynamics is described by the Schr\"odinger equation
\begin{equation}
 i\hbar \frac{\phil}{\partial t}= \Big( - \frac{\hbar^2}{2m} \frac{\partial^2}{\partial z^2}  + V_\parallel(z) \,\Big)\, \phil    -   \frac{\hbar\Omega_{\rf}}{2}\,\phibA \, .  \label{Eq.1DeffAL}
 \end{equation}

For dilute atom laser beams, the intra-laser interactions can be added in a perturbative way in the expression~(\ref{mueff}). In the guide, it corresponds to the 1D mean-field regime ($a \nl \ll 1$). There $\mueff \sim \hbar\omega_\perp (1+2a \nl) $ and the propagation of $\phil$ along the guide (with the eventual presence of an obstacle $V_{\rm{obst}}$) is governed by the nonlinear equation
\begin{equation}
 i\hbar \frac{\phil}{\partial t}= \Big( - \frac{\hbar^2}{2m} \frac{\partial^2}{\partial z^2}  + V_{\rm{obst}} + 2 \hbar\omega_\perp a\nl \,\Big)\, \phil\; .  \label{Eq.NLSE}
 \end{equation}


\section{\label{App.ALwave} Stationary longitudinal atom laser wave function $\phi_{\rm{L},E}$}

We calculate here the atom laser wave function $\phi_{\rm{L},E}$ at the energy $E=\tilde{\mu}_{\BEC}-\hbar \omegarf$. Those are the stationary solutions of the 1D Schr\"odinger equation in the effective potential $V_\parallel(z)$ [see equation~(\ref{Eq.1DeffAL})]. In dimensionless unity (length and energy units being respectively $\sigma_z$ and $\hbar\omega_z/2$), it writes
\begin{equation}
\left[ - \frac{d^2}{d\zbar^2}\;- \;\zbar^2\, \Theta (1-|\zbar|/\Rzbar)\,+\,\dbar \right] \,\phi (\zbar)=0 \; ,
\end{equation}
where $\dbar= 2\omegarf/\omega_z$ is the frequency detuning and $\Rzbar=\tilde{R}_z/\sigma_z$ the corrected TF radius. In the BEC zone ($\zbar<\Rzbar$), the potential is a inverted harmonic oscillator and the odd and even stationary solutions are \cite{Morse,Fertig:1987}
\begin{eqnarray}
\phi_{\rm{o}}(\zbar) & = & \zbar e^{-i\zbar^2/2} \, \rm{F}(
\,\frac{3-i\dbar}{4}\, | \,\frac{3}{2}\, | \,i\zbar^2\, ) \, ,\\
\phi_{\rm{e}}(\zbar) & = & e^{-i\zbar^2/2} \, \rm{F}( \,\frac{1-i\dbar}{4}\, | \,\frac{1}{2}\, | \,i\zbar^2\, ) \, ,
\end{eqnarray}
where $\rm{F}(a|b|x)$ are the confluent hypergeometric
functions \cite{abramowitz}. The BEC wave function being symmetric, the odd contribution vanishes in the overlap integral of the Fermi golden rule [equation~(\ref{Eq.GoldenRule})]. We thus only consider the even function ($\phi\, (\zbar\leq\Rzbar) = A\, \phi_{\rm{e}} (z)$, with $A$ a prefactor to be determined below) and restrict the analysis to $\zbar \geq0$.

\begin{figure}[t]
\centering
\includegraphics[width=16cm]{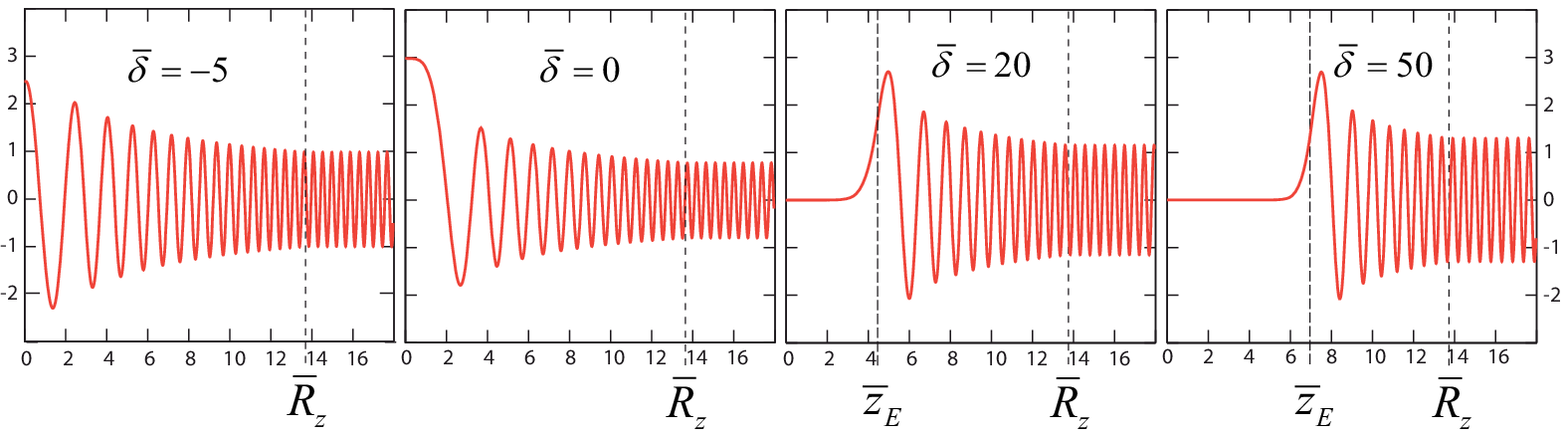}
\caption{Longitudinal wave function $\phi (\zbar)$ for different normalized rf detunings ($\dbar= -$5, 0, 20 and 50). The numerical parameters ($\bar{\mu}_{\BEC}=2\tilde{\mu}_{\BEC}/\hbar\omega_z = 185$, corresponding to $\bar{R}_z=13.6$) are those of section~\ref{par.expcouplage}. $\zbar_E=\dbar^{1/2}$ is the classical turning point position in normalized unit.}
\label{fig.phi}
\end{figure}

Outside the BEC, the wave function $\phi$ is connected to a
stationary plane wave. Introducing a quantization box of size $L$, it takes the form
\begin{equation}
\phi (\zbar>\Rzbar) = \frac{1}{\sqrt{\bar{L} }}\;\sin(\bar{k} \zbar+\varphi) \; ,
\end{equation}
with the dimensionless momentum $\bar{k} =\bar{E}^{1/2}$ associated to the atom laser kinetic energy in the guide.
From the continuity of $\phi$ and its derivative at the BEC edge ($\zbar=\Rzbar$), the two free parameters $A$ and $\varphi$ write
\begin{eqnarray}
A &= &\frac{e^{i\Rzbar^2/2} }{ \sqrt{\bar{L} }} \, \frac{ \sin(\bar{k}\Rzbar+\varphi)}{ \rm{F}(\alpha | 1/2 | i\Rzbar^2)}, \\
\varphi &= &-\bar{k}\Rzbar+ \arctan \left[ \frac{i\bar{k}/\Rzbar}{
1-4\alpha \frac{\rm{F}(\alpha+1 | 3/2 |
i\Rzbar^2)}{\rm{F}(\alpha | 1/2 | i\Rzbar^2)}}\right] .
\end{eqnarray}
Here $\alpha =(1-i\dbar)/4$ and we use the
relation $\frac{d\rm{F}}{dx}(a|b|x) = (a/b)
\rm{F}(a+1|b+1|x)$ \cite{abramowitz}. Note that a WKB calculation will lead to a similar result \cite{Messiah}, apart around the classical turning point $z_E$.

From the knowledge of $\phi_{\rm{L},E}$ (see figure~\ref{fig.phi}) and $\phibA$ [equation~(\ref{Eq.nBtrue})], we calculate the overlap integral in the outcoupling rate expression~(\ref{Eq.GoldenRule}). Here
the density of states for a one-dimensional box of size $L$ is $\eta(E)=L (2m/E)^{1/2}/2\pi\hbar $ (the presence of the mean-field potential in the overlap region with the BEC has a negligible effect for $L\rightarrow \infty$).


\section*{References}

\begin{thebibliography}{42}
\bibitem{Mewes:1997} Mewes M O, Andrews M R, Kurn D M, Durfee D S, Townsend C G and Ketterle W 1997 {\it \PRL} {\bf 78} 582--5
\bibitem{Hagley:1999} Hagley E W, Deng L, Kozuma M, Wen J, Helmerson K, Rolston S L and Philips W D 1997 {\it Science} {\bf 283} 1706--9
\bibitem{Bloch:1999} Bloch I, Hänsch T W and Esslinger T 1999 {\it \PRL} {\bf 82} 3008--11
\bibitem{Cennini:2003} Cennini G, Ritt G, Geckeler C and Weitz M 2003 {\it \PRL} {\bf 91} 240408
\bibitem{nous:GAL} Guerin W, Riou J F, Gaebler J P, Josse V, Bouyer P, Aspect A 2006 {\it \PRL} {\bf 97} 200402
\bibitem{Couvert:2008} Couvert A, Jeppesen M, Kawalec T, Reinaudi G, Mathevet R and Guéry-Odelin D 2009 {\it Europhys. Lett.} {\bf 83} 50001 and 2008 {\it Europhys. Lett.} {\bf 85} 19901
\bibitem{Kleine:2010} Kleine Büning G, Will J, Ertmer J W, Klempt J and Arlt J 2010 {\it App. Phys.} B {\bf 100} 117--23
\bibitem{Carusotto:2001} Carusotto I 2001 {\it \PR} A {\bf 63} 023610
\bibitem{Leboeuf:2001} Leboeuf P and Pavloff N 2001 {\it \PR} A {\bf 64} 033602
\bibitem{Carusotto:2002} Carusotto I 2002 {\it \PR} A {\bf 65} 053611
\bibitem{Paul:2005_1} Paul T, Richter K and Schlagheck P 2005 {\it \PRL} {\bf 94} 020404
\bibitem{Paul:2009PRA} Paul T, Albert M, Schlagheck P, Leboeuf P and Pavloff N 2009 {\it \PR} A {\bf 80} 033615
\bibitem{Gattobigio:2009} Gattobigio G L, Couvert A, Georgeot B and Guéry-Odelin D 2010 {\it \NJP} {\bf 12} 085013
\bibitem{Debs:2010} Debs J E, Döring D, Altin P A, Figl C, Dugué J, Jeppesen M, Schultz J T, Robins N P and Close J D 2010 {\it \PR} A {\bf 81} 013618
\bibitem{Moy:1999} Moy G M, Hope J J and Savage C M 1999 {\it \PR} A {\bf 59} 667--75
\bibitem{Jack:1999} Jack M W, Naraschewski M, Collet M J and Walls D F 1999 {\it \PR} A {\bf 59} 2962--73
\bibitem{gerbier:2001} Gerbier F, Bouyer P and Aspect A 2001 {\it \PRL} {\bf 86} 4729--32, and 2004 {\it \PRL} {\bf 93} 059905(E)
\bibitem{Hope:2000} Hope J J, Moy G M, Collet M J and Savage C M 2000 {\it \PR} A {\bf 61} 023603
\bibitem{Jeffers:2000} Jeffers J, Horak P, Barnett S M and Radmore P M 2000 {\it \PR} A {\bf 62} 043602
\bibitem{Robins:2005} Robins N P, Morison A K, Hope J J and Close J D 2005 {\it \PR} A {\bf 72} 031606(R)
\bibitem{Robins:2006} Robins N P, Figl C, Haine S A, Morrison A K, Jeppesen M, Hope J J and Close J D 2006 {\it \PRL} {\bf 96} 140403
\bibitem{Dalfovo:1999} Dalfovo F, Giorgini S, Pitaevskii L P and Stringari S 1999 {\it \RMP} {\bf 71} 463--512
\bibitem{Burke:1997} Burke J P, Bohn J L, Esry B D and Greene C H 1997 {\it \PR} A {\bf 55} R2511--14
\bibitem{Jackson:1998} Jackson A D, Kavoulakis G M and Pethick C J 1998 {\it \PR} A {\bf 58} 2417--22
\bibitem{Gardiner} Gardiner C W and Zoller P 1991 {\it Quantum Noise} (Berlin, Heidelberg: Springer)
\bibitem{Band:1999} Band Y B, Julienne P S and Trippenbach M 1999 {\it \PR} A {\bf 59} 3823--31
\bibitem{gerbier:thesis} Gerbier F 2003 {\it PhD thesis} Université Paris VI
\bibitem{CohenAnglais:1992} Cohen-Tannoudji C, Dupont-Roc J and Grynberg G 1992 {\it Atom-Photon interactions: Basic Processes and Applications} (New York: John Wiley and Sons)
\bibitem{nous:RSI} Fauquembergue M, Riou J F, Guerin W, Rangwala S, Moron F, Villing A, Le~Coq Y, Bouyer P, Aspect A and Lécrivain M 2005 {\it \RSI} {\bf 76} 103104
\bibitem{Desruelle:1999a} Desruelle B, Boyer V, Murdoch S G, Delannoy G, Bouyer P, Aspect A and Lécrivain M 1999 {\it \PR} A {\bf 60} R1759--62
\bibitem{Stenholm:1997} Stenholm S and Paloviita A 1997 {\it J. Mod. Opt.} {\bf 44} 2533--50
\bibitem{Paloviita:1997} Paloviita A, Suominen K A and Stenholm S 1997 {\it \jpb} {\bf 30} 2623--32
\bibitem{Zobay:2004} Zobay O and Garraway B M 2004 {\it \PR} A {\bf 69} 023605
\bibitem{Bernard:thesis} Bernard A 2010 {\it PhD thesis} Université Paris VI
\bibitem{Billy:2008} Billy J, Josse V, Zuo Z, Bernard A, Hambrecht B, Lugan P, Clément D, Sanchez-Palencia L, Bouyer P and Aspect A 2008 {\it Nature} {\bf 453} 891--4
\bibitem{Kozuma:1999} Kozuma M, Deng L, Hagley E W, Wen L, Lutwak R, Helmerson K, Rolston S and Philips W D 1999 {\it \PRL} {\bf 82} 871--5
\bibitem{Duan:2010} Duan Z, Fan B, Yuan C H, Cheng J, Zhu S and Zhang W 2010 {\it \PR} A {\bf 81} 055602
\bibitem{Gerbier:2004} Gerbier F 2004 {\it Europhys. Lett.} {\bf 66} 771777
\bibitem{Morse} Morse P M, Feshbach H {\it Methods of Theoritical Physics} (New York: McGraw-Hill)
\bibitem{Fertig:1987} Fertig H A and Halperin B I 1987 {\it \PR} B {\bf 36} 7969--76
\bibitem{abramowitz} Abramowitz M and Stegun I A 1972 {\it Handbook of Mathematical Functions} (New York: Dover)
\bibitem{Messiah} Messiah A 1958 {\it Quantum Mechanics} (New York: John Wiley and Sons)
\end{thebibliography}
\end{document}